# A Framework for Scientific Paper Retrieval and Recommender Systems

**Technical Report**

Aravind Sesagiri Raamkumar, Schubert Foo, Natalie Pang

Wee Kim Wee School of Communication and Information,
Nanyang Technological University, Singapore
{aravind002, sfoo, nlspang}@ntu.edu.sg

**Abstract.** Information retrieval (IR) and recommender systems (RS) have been employed for addressing search tasks executed during literature review and the overall scholarly communication lifecycle. Majority of the studies have concentrated on algorithm design for improving the accuracy and usefulness of these systems. Contextual elements related to the scholarly tasks have been largely ignored. In this paper, we propose a framework called the Scientific Paper Recommender and Retrieval Framework (SPRRF) that combines aspects of user role modeling and user-interface features with IR/RS components. The framework is based on eight emergent themes identified from participants feedback in a user evaluation study conducted with a prototype assistive system. 119 researchers participated in the study for evaluating the prototype system that provides recommendations for two literature review and one manuscript writing tasks. This holistic framework is meant to guide future studies in this area.

## 1 Introduction

The quality of a research study and the subsequent publication of research results is inherently associated with the quality of literature review (LR) performed at the start of the study [1]. Special purpose information retrieval (IR) and recommender systems (RS) implementations have been devised for providing relevant research papers to different LR search and manuscript writing (MW) tasks [2, 3]. Two issues are observed in such implementations: Firstly, the applications are piecemeal approaches thereby forcing the researcher to depend on multiple systems to complete important LR search tasks. Secondly, there are a wide variety of algorithms and data items used in these studies, making it a difficult proposition for a contextual integration of services. With the aim of addressing these issues, we selected two key LR search tasks and one MW task for developing a system called Rec4LRW [4]. The three tasks offered by the system are (i) building a reading list of research papers, (ii) finding similar papers based on a set of papers, and (iii) shortlisting papers from the final reading list for inclusion in manuscript based on article type preference of the user [5]. The recommendation techniques of the tasks are based on a set of features that capture the important characteristics of the research paper and the constituent bibliographic references and citations. Along with the traditional metadata fields displayed with the recommended papers, new informational display features were introduced in the system to help the user in making faster and efficient decisions on the relevance and usefulness of papers.

To evaluate the system, we conducted a user evaluation study with 119 researchers who had experience in writing research papers. Both quantitative and qualitative evaluation data were collected. The quantitative section comprised of survey-style questions meant for recording user responses on multiple evaluation measures such as usefulness, satisfaction, relevance, etc. The qualitative section was specifically for eliciting subjective feedback on the preferred features of the tasks and the system along with critical comments and overall feedback of the system.

In this paper, we first present the emergent themes derived from the feedback comments of the participants. Secondly, these themes are further utilized for conceptualizing a specialized framework called scientific paper retrieval and recommender Framework (SPRRF). This framework is meant to guide our future studies with the Rec4LRW system and also to help researchers in better designing systems meant for recommending papers. The framework is intended to be useful in studies where recommendation techniques are conceptualized for scholarly search tasks. Such studies tend to follow a siloed approach with aspects from IR/RS solely considered. SPRRF integrates elements from user modeling [6], IR/RS, search user interfaces (SUI) [7] and exploratory search [8]; therefore most of the contextual entities related to a task are reinforced to complement each other.

We position this new framework in the series of studies conducted with emphasis on combining aspects from tangentially related disciplines. McNee *et al.* [9] considered elements of information seeking models and human-computer interaction (HCI) for proposing the human-recommender interaction (HRI) framework. Champiri *et al.* [10] used "context" as a frame of reference to synthesize RS studies for identifying gaps and opportunity areas. Wolfram [11] raised the case for incorporating language models from natural language processing (NLP) research in bibliometric-enhanced IR studies. Such an approach is aimed at achieving enhanced retrieval and also in understanding boundaries between academic disciplines.

Even though, the frameworks from these earlier studies provide valuable insights, the conceptualization process is predominantly not grounded on empirical user data. The components of SPRRF have been construed based on the feedback data from a large-scale user evaluation study conducted with 119 participants. Therefore, the generalizations observed from the users' responses, can be expected to be positively validated by future research.

This paper is organized as follows. The Rec4LRW prototype system is introduced in section two. Details about the user evaluation study are provided in section three. In section four, the emergent themes derived from the participants' subjective feedback, are presented. The resultant SPRRF framework is introduced in section five and the concluding remarks are outlined in the final section.

## 2    Prototype System

Among the three tasks offered in the Rec4LRW system [4], Tasks 1 and 2 can be considered as two of the important LR search tasks. The third task was aimed to help researchers in preparatory stages of manuscript writing. In the area of manuscript writing, techniques have been proposed to recommend articles for citation contexts in manuscripts [2, 12]. An unexplored area is helping researchers in identifying important and unique papers that are potential candidates for citation in the manuscript. This identification of papers is seen to be affected by the type of article that the author is intending to write, an area that can be explored.  The tasks in the Rec4LRW system are interconnected using two paper collections namely *seed basket* and *reading list*. The seed basket is used for selecting papers from Task 1, and these papers are used as input for Task 2. Reading list is a running collection of papers from both Task 1 and 2, meant for usage in Task 3. Whenever a paper is added in the seed basket, it also gets added to the reading list.

In the task screens of the Rec4LRW system, new informational display features are included for helping researchers in understanding the uniqueness of the recommended papers. For all the three tasks, information cue labels depicting the paper-type of the recommended paper are displayed. The four labels used are *popular, high reach, survey/review and recent*. In Task 1, a word cloud generated with the author-specified keywords of the recommended papers, is embedded in the screen. In Task 2, two shared co-relations features are displayed with each of the recommended papers. These are shared co-references and co-citations of the recommended papers with the papers from the seed basket. In Task 3, a feature to view papers in the parent cluster is provided for each shortlisted paper since the task employs a community detection algorithm to identify clusters with the citation network of the researcher's reading list.

An extract from the ACM Digital Library (ACM DL) is used as the corpus of the system. The extract comprises of papers from proceedings and periodicals, for the period from 1951 to 2011. The sample set for the evaluation study was formed by extracting papers with full text and metadata availability in the extract. The final corpus contained a total of 103,739 articles.

## 3  User Evaluation Study

A user evaluation study was conducted to determine the usefulness and efficiency levels of the three recommendation tasks and the overall system, in the context of same real-world LR and MW tasks. Researchers with experience in writing research papers were recruited for the study. An online pre-screening survey was conducted to screen the potential participants. The study was conducted between November 2015 and January 2016. A total of 119 participants completed the whole study and performed the evaluation of the three tasks and the overall system.

A user guide[1] with the necessary instructions was provided to the participants at the start of the study. In Task 1, the participants had to select a research topic from a list of 43 research topics. On selection of topic, the system provided 20 recommendations. In Task 2, they had to select a minimum of five papers from Task 1 in order for the system to retrieve topically similar papers. For the third task, the participants were requested to include at least 30 papers in the reading list. The minimum paper count in the reading list was set to 30 as the threshold for highest number of shortlisted papers was 26 (for the article-type 'generic research paper'). The three other article-types provided for the study were *case study, conference full paper* and *conference poster*. The shortlisted papers count for these article-types was fixed by taking average values from the references count of the related papers from the ACM DL extract. The participants had to select the article-type and run the task. The evaluation screen in each task was accommodated at the bottom of the screen. The participants had to answer survey questions and subjective feedback questions as a part of the evaluation.

Participants' subjective feedback responses were coded by the corresponding author using an inductive coding style [13]. The aim of the coding exercise was identifying the central themes from the comments.

---

[1] Rec4LRW system user guide can be accessed at http://155.69.254.57/rec4lrw/user_guide.pdf

## 4    Emergent Themes

Since each comment from the participants was assigned with a primary code and an optional secondary code, the formation of themes from these codes became apparent as the coding exercise progressed towards completion. In this section, eight major themes are presented along with few corresponding user comments. The current state of research for each theme is discussed.

### 4.1    Distinct User Groups

Information Systems (IS) across different domains, provide content based on the specific role of the user [14]. The role can determine both the display features and the content to be displayed to the user. In industrial and corporate IS, these roles are utilized to enforce security settings simulating the hierarchy of employees. In academic digital libraries, these roles have not been considered extensively even though attempts have been made to classify users based on varying experience levels [9]. This type of classification can be challenged in relation to the task. In the tangential area of Technology Enhanced Learning (TEL) RS, coursework materials have been recommended to students based on the levels of proficiency [15]. However, research papers provide very little scope to be similarly recommended based on experience level. Perhaps, books can be classified from beginner to expert level and used for recommendations. Conversely, research papers can be classified on content-oriented aspects such as quality of research, extent of contribution, article-type and parent discipline.

From the participants' feedback, the existence of two user groups was inherently visible. One group required control features in the UI for sorting the recommendations and viewing the articles through topical facets. These participants also gave preferences on the algorithm for retrieving papers as researchers tend to follow distinctive paths to arrive at the required papers. The below comments are representative for such users.

> *"..Maybe a side window with categories like high reach, survey etc could be put up and upon clicking it, more papers in that category could be loaded."*
> *"It would be much better if you follow the embedded system classifications, suggested by IEEE/ACM and pick up a relevant paper from each class…"*

The other group of users was largely satisfied with both the recommendations quality and the ranked display of papers. They were not interested in manipulating the display for achieving alternative rankings. Secondly, they trusted the background algorithms used for the recommendations.

*"I hope this becomes an actual web based system that researchers can use."*
*"The idea of providing this system is quite\* good. Such a system if developed and prepared well, can help and speed up the process of literature survey by helping to find better papers…"*

### 4.2 Information Cues

The utility of information cues in positively impacting users' perceptions has been underlined in earlier studies [16]. The usage of information cue labels is new to academic digital libraries although its effectiveness has been proved in other domains [17]. Rec4LRW's unique informational display features such as the information cue labels is an example of cues that enabled the participants in better understanding the recommended papers. Some of the representative comments from the participants are as follows.

*"I like the highlighted recommendations - for e.g. Popular, Recent etc. which greatly helps in distinguishing various references and catches the eye !"*
*"Ease of determining whether the papers were popular/recent/high reach (based on colour coding)"*

Apart from the four cue labels from the current design of the Rec4LRW system, more labels highlighting the unique aspects of the recommended papers can be introduced. In Table 1, the new labels and the corresponding descriptions are proposed. We are encouraging the use of such cue labels as most of the participants felt that these labels were useful during evaluation.

| Label Name | Label Description |
|---|---|
| Interdisciplinary | This label is displayed if the bibliographic references of the papers are from "far-apart" disciplines |
| Tier-1 | This label is displayed if the paper is published in a tier-1 venue |
| Tier-2 | This label is displayed if the paper is published in a tier-2 venue |
| Popular (Views) | This label is displayed if the paper has high viewership in the last month (top 5% percentile in the parent research topic) |
| Popular (Downloads) | This label is displayed if the paper has high number of downloads in the last month |
| Popular (Altmetrics) | This label is displayed if the paper has high altmetric score [18] |
| Article-Type | This label highlights the article-type of the paper (for e.g., conference full paper, conference poster, conceptual paper, technical paper etc.) |

**Table 1.** Proposed information cue labels and their descriptions

### 4.3   Forced Serendipity vs Natural Serendipity

Serendipitous discovery of research papers is a challenging problem as it is complex to model the interestingness of particular unread papers to researcher's current interests. This problem has been handled before in earlier studies [19, 20]. The approaches from these studies are to be classified under the *forced serendipity* category as the resultant recommendations are based on corresponding models. The alternate way of serendipitously encountering research papers is based on purely un-modelled scenarios. For instance, the 'View Papers in the Parent Cluster' feature in the Rec4LRW system helped participants in noticing papers which they have not read earlier. The corresponding comments are as follows.

*"The view papers in the parent cluster function is very helpful to get a full picture of research field."*
*"The user can view many papers in the parent cluster in addition to the shortlisted papers. Thus the user need not spend much time on finding related papers."*

In addition, it can be stated that natural serendipity can be facilitated by incorporating more transparency in the recommendation process. For instance, if the users are able to witness the papers at different stages of filtering, they could perceive certain papers as interesting as against the system's filtering logic. This approach will be considered in the future versions of the Rec4LRW system.

### 4.4   Learning Algorithms vs Fixed Algorithms

The recommendation and retrieval algorithms proposed in earlier studies have been predominantly static and fixed. McNee *et al.* [9] had raised the point on algorithm selection but the claim was restricted to selecting algorithms suitable for corresponding evaluation metrics. The obvious advantage of fixed algorithms is the validity and reproducibility. Nonetheless, factors such as relevance feedback-based changes [21] and choice of algorithms are to be considered for futuristic systems. These two factors contribute to the fluidity level in algorithms. In the case of the first factor, user's actions and choices dictate future recommendations. For the second factor, users expect a list of appropriate algorithms to be presented to them. Some participants in the study suggested heuristics to identify papers for Task 1 and 2. The corresponding comments are presented below.

*"..Take a high impact paper (based on citation and may be exact keyword matching), then go through its own references to understand more about the*

*research conducted. This is because, a good work generally cites other prominent works in the field…"*
*"…The other option is to simply browse the papers from top journals/conferences, which have an embedded systems track and report those…"*

Providing a list of algorithms is expensive in terms of computational capability as these algorithms need to be optimized for superior performance. Nevertheless, user satisfaction will probably improve with algorithmic independence. Some of the related user comments are as follows.

### 4.5 Inclusion of Control Features

In digital libraries, the importance of control features in UI cannot be overstated as these systems serve as an entry point to the large corpuses of papers. Even though, algorithms help in ranking the top most relevant papers for a user's search requirement, not all users would want to select the papers from the ranked list. For instance, if the search engine retrieves *N* number of papers matching the search keywords, users rely on control features such as sort options, topical facets and advanced search features for identifying the desired *n* number of papers. During the user evaluation study, it was noticed that many users felt handicapped by the absence of control features in the Rec4LRW system. This was a surprising observation since the Rec4LRW system was projected as a specialized recommender system, thereby differentiating it from the traditional search systems. The corresponding comments are provided below.

*"Really good for the initial review. It would be nice to see additional filters to focus on a specific topic"*
*"More recent papers shall be included, and it is better if the user can sort the recommended paper by sequence such as sort times, date, relevance..."*

There is a type of determinism assumed in RS where the expectation is that the users will merely trust the ranked list of recommended papers. Nevertheless, the comments from some curious participants raise the case for including traditional control features in these specialized systems. With these features included in the user interface, users will enjoy dual benefits. The first benefit is the access to an initial list of recommended papers which are ranked based on a certain set of preset rules. The second benefit is the ability to manipulate the recommendations list with control features so that the path to desired papers is consciously shortened by the users themselves.

### 4.6 Inclusion of Bibliometric Data

In a narrative similar to the previous theme, informational display features in RS mostly do not represent an extensive set of bibliometric data. In traditional digital libraries, the inclusion of this data has become commonplace as users rely on these metrics for relevance judgment. However, in the case of previous RS, only simple metrics such as the citation count and reference count are included. There are exceptional cases where new metrics are introduced (for e.g., Eigenfactor metric [22]). In the user study, participants explicitly stated the need to include metrics such as impact factor and h-index along with the other metadata. The corresponding comments are provided below.

*"Categorizing the papers based on popularity, journal impact factor, and etc"*
*"…In case that an item in the recommendation list is a journal paper, can we also know its impact factor and which databases indexes it?"*

The main challenge for including these metrics in the user interface is the computing overhead for calculating these values for all the papers in the corpus. Further exacerbating this issue, most of the prototype systems use different datasets [23], thereby re-use of metrics data is not a viable option.

### 4.7 Diversification of Corpus

The evaluation of algorithms in most of the prior studies has been restricted to datasets from certain disciplines such as computer science and related disciplines. Even though there is large level of uniformity in hard and soft sciences on the approaches followed for scientific information seeking [24], not much is known about the differences in relevance heuristics for LR tasks. Therefore, future studies should include papers from "far-apart" disciplines for the evaluation. Some of the corresponding comments from the participants of the user study are provided below.

*"…Due to limitation of data sets (as only ACM papers) search result is not of decent quality."*
*"But in general the main drawback is that "the papers in the corpus/dataset are from an extract of papers from ACM DL". As I work at the intersection of information systems and business many relevant papers are not included in the list."*

### 4.8 Task Interconnectivity

In systems where multiple search tasks are supported, task interconnectivity mechanism is an essential component. With this component, certain redundant user actions can be avoided. In the user study, a good number of participants appreciated the utility of seed basket and reading list towards management of the paper across the three tasks. Some of the corresponding user comments are provided below.

*"I like the idea of giving recommendations based on a seed group of articles, but there needs to be more facets to select from, there needs to be greater selection of seeding articles as well in terms of those facets."*
*"The whole idea seems good for me, especially making seed of 5+ for expanding the bunch."*

## 5 The Framework

On the similar lines of human recommender interaction (HRI) theory [9], we propose the *Scientific Paper Retrieval and Recommender Framework (SPRRF)*, a specialized framework meant to cater for assistive systems of this domain. There are three high-level components in this framework – User Roles, System Customization and User Personalization. The framework is illustrated in Figure 1.

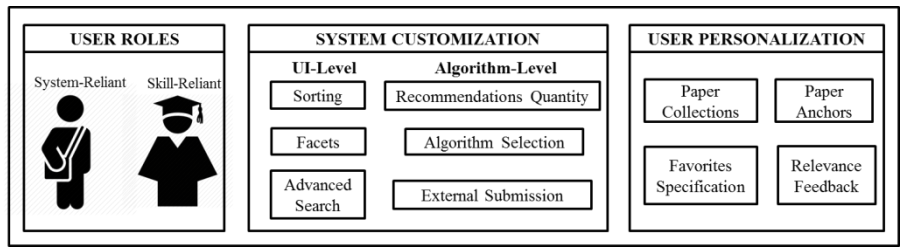

**Fig. 1.** Scientific Paper Retrieval and Recommender Framework (SPRRF)

### 5.1 User Roles

Our proposed classification of user-role is based on the levels of system customization and user personalization preferred by the user. The two proposed roles are (i) skill-reliant and (ii) system-reliant users, corresponding to the theme in Section 4.1. Skill-reliant user role caters to users who prefer to customize the UI and system-level features to a high extent. These users prefer to have sorting, advanced search and filtering options to sieve through the recommended results. They prefer to have control over algorithm logic and the required quantity of output papers. On the other hand, System-reliant role caters to users who prefer to trust the system in its default settings i.e. fixed algorithm logic, low levels of customization and non-personalized options. Research students mostly fall under this category since they preferred the system to an higher extent [4]. The characteristics of these user roles are influenced by the other two components in SPRRF. Assignment of the role to a new user to the system can be made with a simple selection from the user during the first visit. Accordingly, a user could change the role setting during future visits. In a realistic scenario, the user changes the role from skill-reliant to system-reliant when there is a need to further analyze the recommended papers.

### 5.2 System Customization

System Customization component is meant to integrate aspects from UI and algorithms (both IR and RS) for providing a holistic user experience. Earlier studies have paid less attention to this type of integration. Under this component, there are two sub-components. These are UI and algorithm customizations.

**UI Customization.** This sub-component involves control features, corresponding to the fifth theme in Section 4.5. Although, there are different types of control features, three main features are considered adequate. These are (i) sort options, (ii) topical facets and (iii) advanced search options. Sort options provide alternative schemes such as sorting by publication date, citation count and textual similarity. Topical facets are hyperlinks provided in the navigation pane of the results page. The author-specified keywords from research papers are ideal candidates for topical facets. Advanced search options include more text boxes for executing field-specific search queries which can be combined using Boolean operators. These UI customization features in specialized RS will help in simulating a familiar experience for users who have been using traditional digital libraries.

**Algorithmic Customization.** The second level of system customization is related to the retrieval/recommender algorithm. There are three customization features. These are (i) setting the recommendations count, (ii) selecting the algorithm and (iii) submission of external papers through Bibtex files. In all the previous studies and the current study, the recommendations count has been fixed by the researchers based on different rationale. Nevertheless, users will be benefited with this flexible option of setting recommendations count. On a down side, papers with very low relevance scores could be retrieved if the recommendations count is set high. In correspondence to the fourth theme in Section 4.4, certain tasks such as the Task 1 in the current study provide scope for choosing from different algorithms. These algorithms use different rules and information paths for identifying the candidate papers. Hence, the available algorithms could be provided as choices to users for selection. The third feature is the 'upload' option for loading Bibtex files (as in theadvisor tool

[3]). With this feature, users can upload the seed papers into the system using the Bibtex format.

## 5.3 User Personalization

The extent of user personalization applicable for scientific paper recommendations is limited in comparison with other domains such as e-commerce, films and music. Personalization has been largely limited to recommendations based on researcher's publication history [25] and query logs [26]. Through the SPRRF, a different perspective of personalization is presented with four features. These are (i) paper collections, (ii) favorites specification, (iii) paper anchors and (iv) relevance feedback. The seed basket and reading list which are already available in the Rec4LRW system are apt paper collection features for enforcing explicit personalization at task level. This feature addresses the eighth theme in Section 4.9. Anchoring or pinning certain papers in the seed basket or reading list, is the second feature meant for exerting strong influence on recommendations. This helps in acquiring highly personalized results. Alternatively, different weights could be set to the seed papers so that recommendations could be formulated accordingly. User specification of favorites among authors, conferences and journals is the third personalization feature for manipulating recommendations. This feature is set at the user profile level, thereby making these favorites global for all the recommendation tasks carried out by the user. Relevance feedback based re-orientation of recommendations is the fourth feature of user personalization that can really benefit researchers in training the system to their individual tastes. This feature corresponds to fifth theme in section 4.4 where the point of learning algorithms was used.

In Table 2, we have tried the three components of the SPRRF so that systems could implement the framework. However, the mapping has been subjectively performed. We would be validating the framework and the mapping in our future studies.

| SPRRF Feature | Skill-Reliant User | System-Reliant User |
|---|---|---|
| *UI Customization* | | |
| Sort options | √ | |
| Topical Facets | √ | √ |
| Advanced search options | √ | |
| | | |
| *Algorithmic Customization* | | |
| Setting the recommendations count | √ | √ |
| Selecting the retrieval | √ | |
| algorithm | | |
| Submitting external papers | √ | √ |
| *User Personalization* | | |
| Paper collections | √ | √ |
| Favourites specification | √ | √ |
| Paper anchors | √ | |
| Relevance feedback | √ | |

**Table 2.** Mapping the SPRRF components

## 6   Conclusion

In this paper, we have proposed a specialized framework meant to cater for future studies in task-based scientific paper retrieval and recommender systems. The framework has been generated based on eight prominent themes inferred from users' feedback data. The data was collected through a user evaluation study conducted with Rec4LRW system, a prototype built for assisting researchers in three LR and MW tasks. The framework differentiates user roles by system customization and user personalization features.  The unique characteristic of this framework is that it contextually incorporates aspects of IR/RS, exploratory search and user role modeling in an environment where multiple scholarly tasks are handled. As a part of future work, we plan to use the framework to guide the development of next versions of Rec4LRW system.

**Acknowledgements.** This research is supported by the National Research Foundation, Prime Minister's Office, Singapore under its International Research Centres in Singapore Funding Initiative and administered by the Interactive Digital Media Programme Office.